\documentclass[12pt]{article}
\usepackage{amsfonts,latexsym,graphicx,amsmath,cite}

\usepackage{comment} 
\usepackage[usenames,dvipsnames]{color}
\usepackage{amssymb,amsmath,amscd}
\usepackage{dsfont}
\usepackage{amsthm}
\usepackage{multirow}
\usepackage{verbatim}   % useful for program listings and comments
\usepackage{amsfonts}     % complex real number fonts
 \usepackage{pifont} % allows ticks
\usepackage[color,all,cmtip]{xy}
\newcommand{\ketab}[1]{|#1\rangle}
\newcommand{\eeaa}[1]{\begin{align}#1\end{align}}
\newcommand{\eeaan}[1]{\begin{align*}#1\end{align*}}

\def\ba{\begin{array}}
\def\ea{\end{array}}

\makeatletter
\renewcommand{\paragraph}{\@startsection{paragraph}{4}{0ex}%
   {-3.25ex plus -1ex minus -0.2ex}%
   {1.5ex plus 0.2ex}%
   {\normalfont\normalsize\bfseries}}
\makeatother

\theoremstyle{plain}

\begin{document}

\title{Exactly Solvable BCS-BEC crossover Hamiltonians}
\author{A. Birrell$^*$, P. S. Isaac, and J. Links\\
School of Mathematics and Physics, The University of Queensland,\\
Brisbane, Queensland 4072, Australia.\\
*E-mail: a.birrell@uq.edu.au
}
\maketitle

\begin{abstract}
We demonstrate a novel approach that allows the determination of very general classes of exactly solvable Hamiltonians via Bethe ansatz methods. This approach combines aspects of both the co-ordinate Bethe ansatz and algebraic Bethe ansatz. The eigenfunctions are formulated as factorisable operators acting on a suitable reference state. Yet, we require no prior knowledge of transfer matrices or conserved operators. By taking a variational form for the Hamiltonian and eigenstates we obtain general exact solvability conditions. The procedure is conducted in the framework of Hamiltonians describing the crossover between the low-temperature phenomena of superconductivity, in the Bardeen-Cooper-Schrieffer (BCS) theory, and Bose-Einstein condensation (BEC). 
\end{abstract}
%\keywords{Exactly Solvable Models; Bethe Ansatz; Condensed Matter Theory.}

\section{Introduction}

Recent work suggests an adequate understanding of the nature of the BCS-BEC crossover requires the study of Hamiltonians exhibiting coupling between Cooper pairs of atoms and bosonic molecular modes\cite{hkcw01,og02}. Further, the necessity of exact solutions to properly explain the physics of superconducting devices such as ultrasmall superconducting grains\cite{sddvb00} motivates the search for exactly solvable generalisations of the BCS theory.  

A standard technique in this field is the implementation of the Quantum Inverse Scattering Method (QISM)\cite{stf79,tf79,ks79} to construct quantum Hamiltonians with multiple conserved operators, and then find their exact solution by Bethe ansatz methods. Typically, a solution of the Yang-Baxter equation (YBE)\cite{m64,y67,b72} is used to construct a {\it transfer matrix} which generates the conserved operators of the Hamiltonian. The exact solution may then be obtained via Bethe ansatz techniques. We note that in principle the implementation of the co-ordinate Bethe ansatz\cite{b31} does not depend on any prior knowledge of an associated solution of the YBE, nor the conserved operators of the Hamiltonian that it generates. A more modern approach in the framework of the QISM is the algebraic Bethe ansatz\cite{stf79,tf79,ks79}.

The strategy demonstrated here enables the direct determination of general classes of exactly solvable Hamiltonians, with both a bosonic and Cooper pairing degrees of freedom, such that they model BCS-BEC crossover behaviour. By contrast with standard techniques, this gives a unified construction for classes of exactly solvable Hamiltonians with multiple free coupling parameters. By taking appropriate limits we recover seven exactly solvable subcases from the literature\cite{r63,g76,dhl06,ilsz09,dilz11,lrdo11,dlrrr11}.
 
In Section \ref{section2} a general Hamiltonian that models the BCS-BEC crossover behaviour is introduced. In Section \ref{sec:varansatz} the method of obtaining exact solvability constraints for the variational Hamiltonian is outlined.  Calculations that have been omitted for brevity appear in the literature\cite{bil12}.   
   
\section{\label{section2}Variational BCS-BEC crossover Hamiltonian}
We consider the general Hermitian family of pairing Hamiltonians coupled to a bosonic degree of freedom
\eeaa{\label{genHAM} H = H_0-H_1}
where
\eeaa{
\label{H0eq} H_0 &= \alpha N_0 + \kappa N_0^2+ \sum_{k=1}^L f(z_k)N_k,\\
\label{Teq} H_1 &=  \beta \sum_{k=1}^L g(z_k)b_0 b_k^\dagger + \beta\sum_{k=1}^L \overline{g(z_k)} b_0^\dagger b_k +\sigma\sum_{k,s}^L g(z_k)\overline{g(z_s)}b_k^\dagger b_s,
}
for some complex-valued function $f(z)$ and real-valued function $g(z),$ and real-valued parameters $\alpha,$ $\kappa,$ $\beta$ and
$\sigma,$ which will be subject to certain solvability constraints yet to be determined. An overline is used to denote complex conjugation. The operators $b_k^\dagger$ and $N_k=b_k^\dagger b_k$ for $k>0$ are hard-core Cooper pair creation and number operators\footnote{The Cooper pair operators are related to the fermionic operators $c_k^\dagger$ via the identity  $b_k^\dagger=c_{-k}^\dagger c_{k}^\dagger$ where fermions are paired such that there is zero total momentum.}. There is a single bosonic mode with operators $b_0^\dagger$ and $N_0$. The particle operators satisfy the following commutation relations:
\eeaan{
\begin{split}[b_j,b_k] =[b_j^\dagger,b_k^\dagger]=0 ~~~\forall~ j,k\geq 0,~~[b_j,b_k^\dagger] = \left\{\begin{array}{cc}I  &~~j,k=0\\ (I-2N_k)\delta_{jk}  & ~~\textrm{else}\end{array}\right.
\end{split}
}
where $\delta_{jk}$ denotes the standard Kronecker delta. The Hamiltonian commutes with the total number operator $N=N_0+\sum_{k} N_k$. Hamiltonians in this family describe a system of bosonic molecules, condensed into a single bosonic degree of freedom, coupled to $L$ Cooper pairs which are bosonic-like pairs of fermions that must observe an exclusion principle. The Hamiltonian consists of a diagonal part $H_0$, given in Eq. \ref{H0eq}, describing the allowed Cooper pair energy levels and the self-interacting bosonic mode, and a cross-interaction part $H_1$, given in equation Eq. \ref{Teq}, describing level dependent molecule-pair coupling and pair-pair couplings. For the special case $\alpha=\beta=\kappa=0$, which suppresses any action of the Hamiltonian on the bosonic part of the underlying Hilbert space, the Hamiltonian Eq. \ref{genHAM} reduces to a general form of the BCS Hamiltonian. It is for this reason we can refer to the general pairing Hamiltonian as a BCS-BEC crossover Hamiltonian.

We will attempt to directly solve the Hamiltonian by formulating the wave-functions as factorisable operators acting on a suitable reference state, analogous to the algebraic Bethe ansatz\cite{stf79}. On the other hand, we note that a co-ordinate Bethe ansatz approach\cite{b31,lw68,s87} need not resort to any prior knowledge of a transfer matrix or a set of conserved operators. We do not expect that the Hamiltonian has an exact solution in general, however, by combining these aspects of the co-ordinate and algebraic Bethe ansatz methods, this technique allows us to obtain exactly solvable models in a very general fashion. In particular, we have found solvability conditions for two sub-classes of the Hamiltonian, namely the cases $i)~\kappa=0$ of no self-interaction term and $ii)~\sigma=0$ of no BCS pair-pair scattering term.

\section{Variational Approach for Exact Solvability\label{sec:varansatz}}

Motivated by the approach of  Richardson\cite{r63}, we assume the ansatz,
\eeaa{
\label{ansatz} \ketab{\Psi} = \prod_{j=1}^M C(y_j)\ketab{0},~~~ C(y) = \gamma(y)b_0^\dagger + \sum_{k=1}^L h(y,z_k)b_k^\dagger,
}
for the eigenstates of Eq. \ref{genHAM}, where $\ketab{0}$ denotes the vacuum state, $h(y,z)$ is yet to be determined, and $y\in\mathbb{C}$. In general, acting the Hamiltonian $H$ on the state $\ketab{\Psi}$ will lead to terms that are linearly independent of $\ketab{\Psi}$:
\eeaan{
H\ketab{\Psi} = E \ketab{\Psi} + \ketab{\Phi}, ~~~\langle\Psi|\Phi\rangle = 0.
}
The objective is to choose appropriate constraints that allow these terms to be isolated. The ansatz will be an eigenstate if the coefficients of the terms linearly independent of $\ketab{\Psi}$ cancel. The Hamiltonian is furnished with an exact solution on the manifold in the coupling parameter space obtained through compatibility of the corresponding constraints. Technical details of the calculation can be found in the literature\cite{bil12}. We introduce the notation
\eeaan{
\ketab{\Psi_j} = \prod_{l\neq j}^M C(y_l)\ketab{0},\hspace{.5in}\ketab{\Psi_{ij}} = \prod_{l\neq i,j}^M C(y_l)\ketab{0}.
}
By direct calculation the action of $H$ on the state $\ketab{\Psi}$ is of the form 
\eeaan{
H\ketab{\Psi} =& \sum_{j=1}^M\sum_{k=0}^L O_1(y_j,z_k)b_k^\dagger\ketab{\Psi_j} + \sum_{j,l\neq j}^M\sum_{k,s=0}^L O_2(y_j,y_l,z_k,z_s)b_k^\dagger b_s^\dagger\ketab{\Psi_{jl}}.
}
for distributions $O_i$ depending on the various functions introduced above. 

In order to determine the exact solution we look to solve the eigenvalue problem $H\ketab{\Psi} = E\ketab{\Psi}$ for some scalar $E$. The set of {\it solvability constraints} required to find such a solution define a manifold in the coupling parameter space along which the Hamiltonian has an exact solution. We briefly remark on the specific choices for the constraints. Imposing a constraint of the form 
\eeaa{
\nonumber O_2(x,y,w,z) = g(w)k(x,y)h(y,z) +g(w)k(y,x)h(x,z),
}
for some function $k(x,y)$ to be determined later, allows the reduction of linear combinations of $b_k^\dagger b_s^\dagger\ketab{\Psi_{ij}}$ and $b_k^\dagger b_0^\dagger\ketab{\Psi_{ij}}$ terms in to linear combinations of $b_k^\dagger \ketab{\Psi_{i}}$ and $b_0^\dagger \ketab{\Psi_{i}}$ terms by use of the definition $C(y_l)\ketab{\Psi_{jl}}=\ketab{\Psi_{j}}$. To keep equations concise we have introduced the parameter $z_0$, however, we take $h(y,z_0)=\gamma(y)$, $g(z_0)=1$. For $\kappa\neq 0 \neq\sigma$, this constraint yields a trivial exact solution due to the incompatibility of the coefficients of $b_k^\dagger b_s^\dagger\ketab{\Psi_{ij}}$ and $b_k^\dagger b_0^\dagger\ketab{\Psi_{ij}}$ terms. Thus the $\kappa= 0$ and $\sigma=0$ cases are treated separately. A constraint is chosen such that the eigenvalues are of a standardised form,
\eeaa{
\nonumber h(y,z) &= \beta^{-1}(\sigma(\alpha-y)+\beta^2)\frac{g(z)\gamma(y)}{f(z)-y},~~~~E=\sum_{j}y_j.
}
Compatibility of these constraints defines $k(x,y)$.
%It can be shown that Hamiltonians of the form Eq. \ref{genHAM} subject to the constraint Eq. \ref{const1} with the aboce energy sepctrum only attain non-trivial exact solutions with the choice Eq. \ref{const2}.
A final constraint is obtained by imposing that the remaining coefficients cancel. These are analogous to the Bethe ansatz equations and, subject to the compatibility of all constraints, the Hamiltonian is exactly solvable if the $y_j$ are their roots. 

For the case $\sigma=0$ compatibility of the constraints leads to the constraining relations for exact solvability:
\eeaan{
 h(y_j,z_k) = \frac{\beta g(z_k)\gamma(y_j)}{f(z_k)-y_j},~~~
 f(z_k) = \kappa^{-1}\beta^2 g(z_k)\overline{g(z_k)} +\kappa^{-1}c_1\\
 y_j-(\alpha+\kappa) +\sum_{k=1}^L\frac{\kappa f(z_k)- c_1}{f(z_k)-y_j} = 2\sum_{l\neq j}^M \frac{c_1-c_2 y_l}{y_j-y_l}, ~~~c_2 =\kappa.
}
for arbitrary constants $\beta$(or $\alpha$), $c_1$, and $c_2$.

In the case $\kappa=0$ the solvability conditions are
\eeaan{
c_1 \sigma\beta = c_2(\sigma\alpha+&\beta^2)\beta ,~~~c_1 = c_2 f(z_k)-\overline{g(z_k)}g(z_k),\\
h(y_j,z_k) = \beta&^{-1}(\sigma(\alpha-y_j)+\beta^2)\frac{g(z_k)\gamma(y_j)}{f(z_k)-y_j},\\
 \frac{y_j-\alpha}{\sigma(\alpha-y_j)+\beta^2}& +\sum_{k=1}^L\frac{c_2 f(z_k)-c_1}{f(z_k)-y_j} =
2\sum_{l\neq j}^M \frac{c_1 -c_2y_l}{y_j -y_l}.
}
for arbitrary constants $\beta$(or $\alpha$), $c_1$, and $c_2$. In either case the $\gamma(y_j)$ will be fixed by any normalisation of $\ketab{\Psi}$.

\section{Final Remarks\label{sec:conclusion}}

\begin{figure}[htbp]
\centering
			$ \xymatrix @!=40pt {&   \sigma=0  \ar[dd]_{c_1=0} &  &   &  \kappa=0 \ar[ddl]_{c_1=0}   \ar[ddr]^{\beta=0} &  & \\
			& & & & & & \\
&  \textrm{BCS-BEC \cite{lrdo11}} \ar[ddl]_{\alpha=\kappa(L-2M)} \ar[ddr]^{\kappa=0} &  &  \textrm{BCS-BEC \cite{dilz11}} \ar[ddl]_{\sigma=0}  \ar[ddr]^{\beta=0} &  &  \textrm{BCS} \ar[ddl]_{c_1=0} \ar[dd]_{\stackrel{c_1<0,}{c_2=-1}}  \ar[ddr]^{c_2=0} &\\
		&	& & & & &\\
 \textrm{BCS-BEC \cite{dhl06}} & &  \textrm{Dicke \cite{g76}}  & &  \textrm{BCS \cite{ilsz09}} & \textrm{BCS \cite{dlrrr11}}  &  \textrm{BCS \cite{r63}} 
} $

\caption{Hierarchy of exactly solvable models that have been recovered. Specific models can be obtained from the references, excluding three previously unknown cases.}
%\caption{Hierarchy of known exactly solvable models that are recovered from the variational approach. Key: The known models are: the 2-channel $p+ip$-wave BCS coupled to a bosonic mode with no pair-pair interaction\cite{lrdo11}\,; the $p+ip$-wave BCS coupled to bosonic mode with no self-interaction term\cite{dilz11}\,; a BCS system coupled to a bosonic mode\cite{dhl06}\,; the Dicke model\cite{g76}\,; the $p+ip$-wave BCS\cite{ilsz09}\,; a BCS model for heavy nuclei\cite{dlrrr11} \,; Richardson's reduced $s$-wave BCS\cite{r63}\,; $[^\dagger]$ is a general BCS model which has not appeared previously.}
	\label{fig:submodels}
\end{figure}
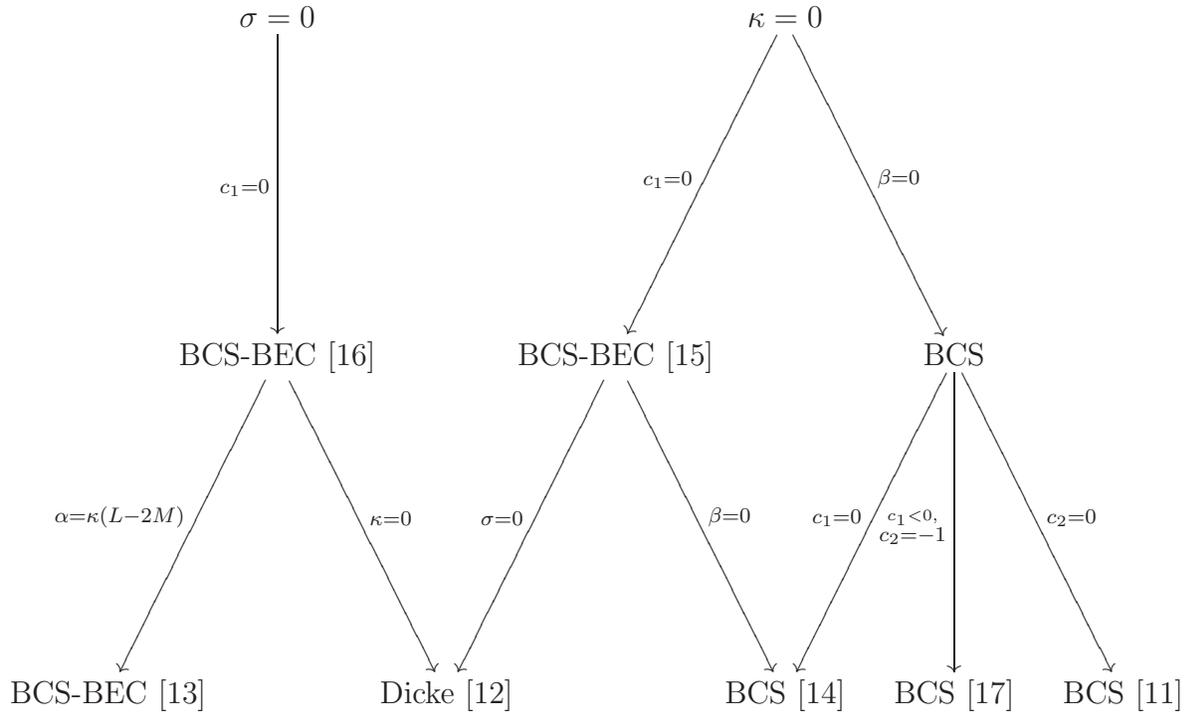

We have determined manifolds in the coupling parameters of Eq. \ref{genHAM} for which an exact solution exists. As shown in Fig. \ref{fig:submodels}, in appropriate limits of the general exact models found above eight subcases are recovered, seven of which are known \cite{r63,lrdo11,dilz11,dhl06,g76,ilsz09,dlrrr11}. Remarkably, this method has enabled a unified construction for exact solutions of models that were previously independent. There is opportunity to extend the scope of this approach to other models\cite{alo01b,des01,ml12} by adopting an ansatz of a different form and relaxing the assumed separability of the pair-pair interaction.

\pagebreak
\subsubsection*{Acknowledgements} This work was supported in part by the Australian Research Council under Discovery Project DP110101414. A. Birrell acknowledges the support of an Australian Postgraduate Award.

\bibliographystyle{plain}
\bibliography{G12_variational_Bib}

\begin{thebibliography}{10}

\bibitem{hkcw01}
M.~Holland, S.~J. J.~F. Kokkelmans, M.~L. Chiofalo and R.~Walser, {\em Phys.
  Rev. Lett.} {\bf 87}, 120406 (2001).

\bibitem{og02}
Y.~Ohashi and A.~Griffin, {\em Phys. Rev. Lett.} {\bf 89}, 130402 (2002).

\bibitem{sddvb00}
G.~Sierra, J.~Dukelsky, G.~G. Dussel, J.~{von Delft} and F.~Braun, {\em Phys.
  Rev. B} {\bf 61}, R11890 (2000).

\bibitem{stf79}
E.~K. Sklyanin, L.~A. Takhtadzhyan and L.~D. Faddeev, {\em Theor. and Math.
  Phys.} {\bf 40}, 688 (1979).

\bibitem{tf79}
L.~A. Takhtadzhyan and L.~D. Faddeev, {\em Russ. Math. Surv.} {\bf 34}, 11
  (1979).

\bibitem{ks79}
P.~P. Kulish and E.~K. Sklyanin, {\em Phys. Lett. A} {\bf 70}, 461 (1979).

\bibitem{m64}
J.~B. McGuire, {\em J. Math. Phys.} {\bf 5}, 622 (1964).

\bibitem{y67}
C.~N. Yang, {\em Phys. Rev. Lett.} {\bf 19}, 1312 (1967).

\bibitem{b72}
R.~J. Baxter, {\em Ann. Phys.} {\bf 70}, 193 (1972).

\bibitem{b31}
H.~Bethe, {\em Z. Phys. A} {\bf 71}, 205 (1931).

\bibitem{r63}
R.~Richardson, {\em Phys. Lett.} {\bf 3}, 277 (1963).

\bibitem{g76}
M.~Gaudin, {\em J. Physique} {\bf 37}, 1087 (1976).

\bibitem{dhl06}
C.~Dunning, K.~E. Hibberd and J.~Links, {\em Nucl. Phys. B} {\bf 748}, 458
  (2006).

\bibitem{ilsz09}
M.~Iba$\tilde{\textrm{n}}$ez, J.~Links, G.~Sierra and S.-Y. Zhao, {\em Phys.
  Rev. B} {\bf 79}, 180501 (2009).

\bibitem{dilz11}
C.~Dunning, P.~S. Isaac, J.~Links and S.-Y. Zhao, {\em Nucl. Phys. B} {\bf
  848}, 372 (2011).

\bibitem{lrdo11}
S.~H. Lerma, S.~M.~A. Rombouts, J.~Dukelsky and G.~Ortiz, {\em Phys. Rev. B}
  {\bf 84}, 100503 (2011).

\bibitem{dlrrr11}
J.~Dukelsky, S.~Lerma, L.~M. Robledo, R.~Rodriguez-Guzman and S.~M.~A.
  Rombouts, {\em Phys. Rev. C} {\bf 84}, 061301 (2011).

\bibitem{bil12}
A.~Birrell, P.~S. Isaac and J.~Links, {\em Inverse Problems} {\bf 28}, 035008
  (2012).

\bibitem{lw68}
E.~H. Lieb and F.~Y. Wu, {\em Phys. Rev. Lett.} {\bf 20}, 1445 (1968).

\bibitem{s87}
P.~Schlottmann, {\em Phys. Rev. B} {\bf 36}, 5177 (1987).

\bibitem{alo01b}
L.~Amico, A.~D. Lorenzo and A.~Osterloh, {\em Nucl. Phys. B} {\bf 614}, 449
  (2001).

\bibitem{des01}
J.~Dukelsky, C.~Esebbag and P.~Schuck, {\em Phys. Rev. Lett.} {\bf 87}, 066403
  (2001).

\bibitem{ml12}
I.~Marquette and J.~Links, {\em Nucl. Phys. B} {\bf 866}, 378  (2013).

\end{thebibliography}

\end{document}